\documentclass[aps,prb,amsmath,twocolumn]{revtex4}
\usepackage{bm}
\usepackage{graphicx}

\begin{document}
\title{Thermal fluctuations in superconducting nanowires}
\author{Dmitri S. Golubev and Andrei D. Zaikin}
\affiliation{Forschungszentrum Karlsruhe, Institut f\"ur Nanotechnologie,
76021, Karlsruhe, Germany}
\affiliation{I.E. Tamm Department of Theoretical Physics, P.N.
Lebedev Physics Institute, 119991 Moscow, Russia}

\begin{abstract}
We re-analyze the problem of thermally activated phase slips
(TAPS) which can dominate the behavior of sufficiently thin
superconducting wires at temperatures close to $T_C$. With the aid
of an effective action approach we evaluate the TAPS rate which
turns out to exceed the rate found by D.E. McCumber and B.I.
Halperin, Phys. Rev. B {\bf 1}, 1054 (1970) within the TDGL
analysis by the factor $\sim (1-T/T_C)^{-1} \gg 1$. Additional
differences in the results of these two approaches arise at bias
currents close to the Ginzburg-Landau critical current where the
TAPS rate becomes bigger. We also derive a simple formula for the
voltage noise across the superconducting wire in terms of the TAPS
rate. Our results can be verified in modern experiments with
superconducting nanowires.
\end{abstract}

\maketitle

\section{Introduction}

Fluctuations are known to play an important role in
superconducting structures with reduced dimensions. In the case of
superconducting nanowires \cite{AGZ} fluctuations may essentially determine
the system behavior in a wide temperature interval causing, e.g.,
non-vanishing wire resistance down to $T=0$.

Over four decades ago it was realized by Little \cite{Little} that
sufficiently thin superconducting wires may acquire a non-zero
resistance below the BCS critical temperature of the bulk material
$T_C$ due to non-trivial thermal fluctuations of the order
parameter $\Delta =|\Delta |\exp (i\varphi )$. Such fluctuations
result in a temporary local destruction of $|\Delta |$ accompanied
by the phase slippage in the corresponding points of the wire. 
According to the Josephson relation $V=\dot \varphi /2e$ this process 
must cause a non-zero voltage drop across the superconducting sample 
thus bringing it into a resistive state.

Quantitative theory of these thermally activated phase slips
(TAPS) was worked out by Langer and Ambegaokar \cite{la} and by
McCumber and Halperin \cite{mh}. This LAMH theory predicts that in
a superconducting wire TAPS are created with the rates
$\Gamma_{\pm}$ which are defined by the activation
dependence
\begin{align}
\Gamma_{\pm }=B_{\pm }e^{-\delta F_{\pm }/T}. \label{Gamma111}
\end{align}
Here $\delta F_{\pm}$ are effective free energy barriers which the
system should overcome in order to create a phase slip corresponding
to the overal phase change $\pm 2\pi$. These potential barriers are essentially
controlled by the superconducting condensation energy for the
volume of the TAPS core where the order parameter $|\Delta |$ gets
destroyed by thermal fluctuations. In the absence of any external
bias one naturally has $\Gamma_{+}=\Gamma_{-}$ and, hence, no net
voltage across the sample can occur. Applying an external current
$I$ one lifts the symmetry between ``positive'' and ``negative'' TAPS. As
a result, there appears a voltage drop (and, hence, non-zero
resistance $R(T)$) proportional to the difference between the two
TAPS rates $\Gamma_{+}-\Gamma_{-}$.

According to Eq. (\ref{Gamma111}) TAPS remain significant only
at temperatures close to $T_C$ while $R(T)$ decreases
exponentially as $T$ is lowered well below the critical
temperature. This prediction was fully confirmed in experiments
\cite{Webb,Tinkham} where the activation behavior of $R(T)$ was
detected in small superconducting whiskers with diameters $\sim
0.5$ $\mu$m. Later is was realized that in thinner wires not only
thermal but also quantum fluctuations of the order parameter
(quantum phase slips) become important. It was demonstrated both
theoretically \cite{ZGOZ,GZ01} and experimentally
\cite{Gio,BT,Lau,Zgirski,Altomare,Bezryadin06,Bezryadin08} that
quantum phase slip effects can yield appreciable resistivity of
superconducting wires with diameters in the range $\sim 10$ nm
even well below $T_C$. For more details we refer the reader to the
review \cite{AGZ}.

Turning again to thermal fluctuations near $T_C$, we note that,
while for evaluation of the free energy barriers for TAPS $\delta F_{\pm }$
in Eq. (\ref{Gamma111}) it suffices just to solve the standard Ginzburg-Landau
(GL) equations, the problem of finding the pre-exponent $B_{\pm}$ is in
general much more involved, as it requires employing the formalism which
properly accounts for {\it dynamical effects} in superconductors.
McCumber and Halperin (MH) \cite{mh} treated this problem within the formalism
of the so-called time dependent Ginzburg-Landau (TDGL) equations \cite{LV} which
was available at that time. Unfortunately, this formalism is known
to suffer from serious drawbacks (see, e.g., Refs. \onlinecite{AGZ,ogzb,LK} for
further discussion) and it is in general hardly applicable below
$T_C$. Thus, although MH calculation \cite{mh} of the pre-exponent $
B_{\pm}$ was correct and sound by itself, their final result needs to be
re-analyzed on the basis of a more solid theoretical approach. This task will
be accomplished below.

The structure of the paper is as follows. In Sec. 2 we briefly
re-capitulate the microscopic effective action formalism \cite{ZGOZ,GZ01,ogzb}
and employ it in order to estimate the fluctuation correction to
the order parameter of ultra-thin superconducting nanowires. In
Sec. 3 we apply this formalism in order to evaluate the pre-exponent $B_{\pm}$
in the expression for the TAPS rates (\ref{Gamma111}). We will then perform a
detailed  comparison between our result and that of Ref. \onlinecite{mh}. In
addition, we will present a simple formula which expresses the voltage noise
in superconducting nanowires via the TAPS rates (\ref{Gamma111}) evaluated here.
Some technical details of our calculation of
fluctuation determinants are relegated to Appendix.

\section{Effective action and Gaussian fluctuations}

Consider a uniform superconducting wire with cross section $s$ and length $X$.
In order to account for superconducting fluctuations in such a wire we will
use the effective action approach developed in Refs.
\onlinecite{ogzb,GZ01}. Our starting point is the path integral representation
of the grand partition function
\begin{align}
{\mathcal{Z}}=\int{\mathcal{D}}\Delta\,{\mathcal{D}}V\,{\mathcal{D}}{\bm A}\;
e^{-S_{\rm eff}},
\end{align}
where $S_{\rm eff}[\Delta , V, {\bm A}]$ is the imaginary time
version of the effective action for a superconducting wire. The
fluctuating order parameter field $\Delta$ as well as the scalar
and vector potentials $V$ and ${\bm A}$ depend on coordinate $x$
along the wire (i.e. $-X/2\leq x\leq X/2$) and imaginary time
$\tau$ restricted to the interval $0\leq\tau\leq 1/T$. The exact
expression for this effective action is obtained by integrating
out the electron degrees of freedom and is not easily tractable in
a general situation. In order to simplify this general expression
for the action $S_{\rm eff}$ one can assume that deviations of the
amplitude of the order parameter field $\Delta(x,\tau)$ from its
equilibrium value $\Delta_{0}$ are relatively small. This
assumption allows to expand the effective action in powers of
$\delta\Delta(x,\tau )=\Delta(x,\tau)-\Delta_{0}$ and in the
electromagnetic fields up to the second order terms. The next step
is to average over disorder. After such averaging the effective
action becomes translationally invariant both in space and in
time. Performing the Fourier transformation we obtain
\cite{ogzb,GZ01}
\begin{align}
S_{\rm eff} & =\frac{s}{2}\int\frac{d\omega dq}{(2\pi)^{2}}\left\{  \frac{|A|^{2}%
}{Ls}+\frac{C|V|^{2}}{s}+ \chi_{D}\left|  qV+\frac{\omega}{c}A\right|
^{2} \right.
\nonumber\\
&  \left. +\,\chi_{J}\left|  V +\frac{i\omega}{2e}\varphi\right| ^{2}
+\frac{\chi_{L}}{4m^{2}}\left| iq\varphi+\frac{2e}{c} A
\right| ^{2} + \chi_{\Delta}|\delta\Delta|^{2} \right\} .\label{a105}%
\end{align}
The functions $\chi_{\Delta}(\omega ,q)$, $\chi_{J}(\omega ,q)$, $\chi_{L}(\omega ,q)$ and
$\chi_{D}(\omega ,q)$ are related to the averaged products of
the Matsubara Green functions. The corresponding general expressions
are established in Refs. \onlinecite{ogzb,GZ01,AGZ}. They are rather
cumbersome and will not be specified here. In what follows we will use
only simplified forms of these functions applicable in certain limits.

As the action $S_{\rm eff}$ (\ref{a105}) is quadratic both in the voltage $V$
and the vector potential $A$, these variable can be integrated out exactly.
After such integration one arrives at the  effective action $S$ which
only depends on $\varphi$ and $\delta\Delta$. We get
\begin{equation}
S=\frac{s}{2}\int\frac{d\omega dq}{(2\pi)^{2}}\left\{ {\mathcal{F}}%
(\omega,q)|\varphi|^{2}+ \chi_{\Delta}|\delta\Delta|^{2}\right\}
.\label{a116}%
\end{equation}
Since usually the wire geometric inductance $L$ remains
unimportant, in what follows we will disregard this quantity by setting
$L=0$. Then we obtain \cite{GZ01,AGZ}
\begin{equation}
\mathcal{F}(\omega,q)= \frac{\left( \frac{\chi_{J}}{4e^{2}}\omega
^{2}+\frac{\chi_{L}}{4m^{2}}q^{2} \right)  \left( \frac{C}{s}+
\chi_{D} q^{2} \right) + \frac{\chi_{J}\chi_{L}}{4m^{2}}q^{2}}
{\frac{C}{s}+\chi_{J}+\chi_{D} q^{2}} ,\label{a106}%
\end{equation}

The effective action (\ref{a116}) allows to directly evaluate the fluctuation
correction to the equilibrium value of the
order parameter in superconducting nanowires. Performing
Gaussian integration over both $\varphi$ and $\delta\Delta$ we arrive at the
wire free energy
\begin{align}
F=F_{BCS} -\frac{T}{2}\sum_{\omega,q}\left[  \ln\frac{\lambda{\mathcal{F}%
}(\omega,q)}{2N_{0}\Delta_{0}^{2}} + \ln\frac{\lambda\chi_{\Delta
}(\omega,q)}{2N_{0}} \right] ,
\end{align}
where $F_{BCS}$ is the standard BCS free energy and
$\lambda$ is the BCS coupling constant. The order
parameter is defined by the saddle point equation $\partial
F/\partial\Delta=0$ and can be written in the form
$\Delta=\Delta_{0}-\delta\Delta_{0}$, where $\Delta_{0}$
is the solution of the BCS self-consistency equation $\partial F_{BCS}%
/\partial\Delta_{0}=0$  and the fluctuation correction
$\delta\Delta_{0}$ has the form
\begin{align}
\delta\Delta_{0}&=-\frac{T}{2}\left(
\frac{\partial^{2}F_{BCS}}{\partial \Delta_{0}^{2}}\right) ^{-1}
\nonumber\\ &\times\,
\frac{\partial}{\partial\Delta_{0}}\sum _{\omega,q} \left[
\ln\frac{\lambda{\mathcal{F}}(\omega,q)}{2N_{0}\Delta _{0}^{2}} +
\ln\frac{\lambda\chi_{\Delta}(\omega,q)}{2N_{0}} \right]
.\label{dDelta}
\end{align}

First let us consider the low temperature limit $T\ll\Delta_0$.
It is useful to note that at large values of the wave number
$|q|\gg \sqrt{\Delta_0/D}$ and/or frequency $|\omega|\gg\Delta_0$
the functions ${\mathcal F}(\omega,q)/\Delta_0^2$ and $\chi_\Delta(\omega,q)$
are weakly affected by superconductivity. Hence, we can restrict the sum in
Eq. (\ref{dDelta}) only to low frequencies $|\omega|\lesssim \Delta_0$ and
wave numbers $|q|\lesssim \sqrt{\Delta_0/D}$.
It will be convenient for us to introduce dimensionless parameters
$y=\omega/\Delta_0$, $z=q\sqrt{D/\Delta_0}$ and
express the kernels as follows
\begin{eqnarray}
\chi_\Delta= N_0 F_\Delta(y,z); &&   \chi_J=e^2N_0F_J(y,z);
\nonumber\\
\chi_L= m^2N_0D\Delta_0 F_L(y,z);&& \chi_D=\frac{e^2N_0D}{\Delta_0}F_D(y,z);
\end{eqnarray}
where all the functions $F_j$ are dimensionless. The function
${\mathcal F}(\omega,q)$ acquires the form
\begin{eqnarray}
{\mathcal F}(y,z)&=&\frac{N_0\Delta_0^2}{4}\bigg[\frac{(y^2F_J+z^2F_L)(C/s+e^2N_0z^2F_D)}
{C/s+e^2N_0F_J+e^2N_0z^2F_D}
\nonumber\\ &&
+\,\frac{e^2N_0 z^2F_JF_L}{C/s+e^2N_0F_J+e^2N_0z^2F_D}
\bigg].
\end{eqnarray}
For a wire of length $X$ we obtain $\partial F/\partial\Delta_0=2N_0sX$ and at $T=0$
the correction to $\Delta_0$ reads
\begin{eqnarray}
\delta\Delta_0 &\simeq& \frac{3}{8} \frac{\Delta_0}{sN_0\sqrt{D\Delta_0}}
\int_{-1}^1\frac{dydz}{(2\pi)^2} \bigg[ \ln\frac{{\mathcal F}(1,1)}{{\mathcal F}(y,z)}
\nonumber\\ &&
+\, \ln\frac{ F_\Delta(1,1)}{F_\Delta(y,z)}  \bigg].
\end{eqnarray}
The integral
$$\int_{-1}^1\frac{dydz}{(2\pi)^2}  \ln\frac{ F_\Delta(y,z)}{F_\Delta(1,1)}$$
is well convergent at small $y$ and $z$,
therefore we can replace it by a constant of order one.
The integral
$$\int_{-1}^1\frac{dydz}{(2\pi)^2}  \ln\frac{{\mathcal F}(y,z)}{{\mathcal
    F}(1,1)}$$
is only slightly
more complicated, since ${\mathcal F}(y,z)\to 0$ for $y,z\to 0$.
However, since the function ${\mathcal F}(y,z)$ enters only under
the logarithm, this integral is convergent as well.
Making use of the above expressions for the functions ${\mathcal{F}}%
(\omega,q)$ and $\chi_{\Delta}(\omega,q)$ at $T
\to 0$ we obtain
\begin{align}
\frac{\delta\Delta_{0}}{\Delta_{0}}\sim \frac{1}{g_{\xi}} \sim
 Gi_{\mathrm{1D}}^{3/2}.\label{estim22}%
\end{align}
Here $g_{\xi}$ is the
dimensionless conductance of the wire segment of length $\xi$ and
$Gi_{\mathrm{1D}}$ is the Ginzburg number for a superconducting nanowire
defined as the value
$(T_{C}-T)/T_{C}$ at which the fluctuation correction to the wire specific heat
becomes equal to the specific heat jump at the phase transition point. In the
case of quasi-1D wires this number reads \cite{LV}:
\begin{align}
Gi_{\mathrm{1D}}=\frac{0.15}{(sN_{0}\sqrt{D\Delta_{0}})^{2/3}}
\label{Gi}
\end{align}

We note that in Eq. (\ref{estim22}) fluctuations of both
the phase and the
absolute value of the order parameter give contributions of the
same order. The estimate (\ref{estim22}) demonstrates that
at low temperatures suppression of the order parameter in
superconducting nanowires due to Gaussian fluctuations remains
weak as long as $g_{\xi} \gg1$ and it becomes important only for
extremely thin wires with $Gi_{\mathrm{1D}} \sim 1$ in which case
the width of the fluctuation region $\delta T$
is comparable to $T_{C}$
and the BCS mean field approach becomes obsolete down to $T=0$.

Turning to higher temperatures we observe that at $T$ sufficiently
close to the critical temperature $T_C$ it is necessary to retain
only the contribution from zero Matsubara frequency.
At the same time the terms originating from all non-zero frequencies are
small in the parameter $\Delta_0(T)/T_C \ll 1$ and, hence, can be
safely omitted. Performing the integration over $q$ we get
\begin{eqnarray}
\frac{\delta\Delta_0}{\Delta_0 (T)}=\frac{T}{\delta F},
\label{dFF}
\end{eqnarray}
where
\begin{align}
\Delta_0(T)=\sqrt{\frac{8\pi^{2}T(T_{C}-T)}{7\zeta(3)}}.
\label{Delta0}
\end{align}
and
\begin{eqnarray}
\delta F=\frac{16\pi^{2}}{21\zeta(3)}
sN_{0}\sqrt{\pi D}(T_{C}-T)^{3/2}
\label{dF0}
\end{eqnarray}
turns out to be exactly equal to the magnitude of the
effective free energy barrier for TAPS in the LAMH theory
\cite{la,mh} in limit of small transport currents (see
below). Eqs. (\ref{dFF}), (\ref{dF0}) demonstrate that
at temperatures close to $T_C$ Gaussian fluctuations of the superconducting
order parameter in thin wires become more significant and effectively
wipe out superconductivity at $\delta F \lesssim T_C$, i.e.
already in much thicker wires than in the case of low temperatures $T \ll T_C$.

\section{Thermally activated phase slips}

In what follows let us restrict our attention to superconducting
wires in which the condition $\delta F_0 \gg T_C$ is well satisfied and, hence,
the effect of Gaussian fluctuations on the order parameter $\Delta_0(T)$
can be safely neglected. This condition requires the wire to be sufficiently
thick and/or the temperature should not be too close to $T_C$,
i.e. $(T_C-T)/T_C \gg Gi_{\rm 1D}$. At the same time we assume that the
temperature is still not far from $T_C$, i.e.$T_C-T \ll T_C$ in which case
the physics is dominated by thermally activated phase
slips \cite{la,mh}. As we already discussed,
sufficiently thin superconducting wires acquire
non-zero resistance even below $T_{C}$ due to TAPS, and this resistance is
essentially determined by the TAPS rates (\ref{Gamma111}).

\subsection{Activation exponent}

The free energy barriers $\delta F_{\pm}$ for TAPS corresponding
to overall phase jumps by $\pm 2\pi$ have been evaluated by Langer
and Ambegaokar \cite{la}. Here we briefly re-capitulate their
results. In order to obtain $\delta F_{\pm }$ entering into
Eq. (\ref{Gamma111}) we make use of the standard Ginzburg-Landau
free energy functional for a wire of length $X$:
\begin{align}
F[\Delta(x)]&= sN_{0}\int_{-X/2}^{X/2} dx \bigg(  \frac{\pi D}{8T}\left| \frac
{\partial\Delta}{\partial x}\right|^{2}
+ \frac{T-T_{C}}{T_{C}} |\Delta|^{2}
\nonumber\\ &
+\, \frac{7\zeta(3)}{16\pi^{2}T^{2}}|\Delta|^{4} \bigg) -\frac{I}{2e}[\varphi(X/2)-\varphi(-X/2)].
\label{FGL}
\end{align}
Here $\varphi(x)$ is the phase of the order parameter $\Delta (x)$ and $I$ is
the external current applied to the wire.

The saddle point paths for this functional are determined by the standard GL
equation
\begin{align}
-\frac{\pi D}{8T}\frac{\partial^{2}\Delta}{\partial x^{2}} + \frac{T-T_{C}%
}{T_{C}} \Delta+ \frac{7\zeta(3)}{8\pi^{2}T^{2}}|\Delta|^{2}\Delta=0.
\label{GL}
\end{align}
For any given value of the bias current
\begin{eqnarray}
I=\frac{\pi eN_0Ds}{2T}|\Delta|^2\nabla\varphi
\end{eqnarray}
this equation has a number of solutions.
The TAPS free energy barrier $\delta F_{+}$ is determined by the
two of them. The first one, $\Delta_m=|\Delta_m|\exp (i\varphi_m)$,
corresponds to a metastable minimum of the free energy functional. This
solution reads
\begin{eqnarray}
|\Delta_{m}|=\Delta_0(T)\sqrt{\frac{1+2\cos\alpha}{3}},\;
\varphi_m= \frac{2T}{\pi eN_0Ds}\frac{Ix}{|\Delta_m|^2}.
\label{Deltam}
\end{eqnarray}
Here $\Delta_0(T)$ is the equilibrium superconducting gap
defined in Eq. (\ref{Delta0}) and the parameter
\begin{eqnarray}
\alpha=\frac{\pi}{3}\theta\left(|I|-\frac{I_C}{\sqrt{2}}\right)+\frac{1}{3}\arctan\frac{2|I|\sqrt{1-(I/I_C)^2}}{I_C\left(1-2(I/I_C)^2\right)}
\end{eqnarray}
accounts for the external bias current $I$. The Ginzburg-Landau
critical current $I_C$ is defined by the standard expression
\begin{eqnarray}
I_C=\frac{16\sqrt{6}\pi^{5/2}}{63\zeta(3)}eN_0\sqrt{D}s(T_C-T)^{3/2}.
\end{eqnarray}

The second, saddle point, solution $\Delta_s(x)=|\Delta_s|\exp (i\varphi_s)$
of Eq. (\ref{GL}) has the form
\begin{eqnarray}
\frac{|\Delta_s|}{\Delta_0(T)}&=&\sqrt{\frac{1+2\cos\alpha}{3}
-\frac{2\cos\alpha-1}{\cosh^2\left[\sqrt{2\cos\alpha-1}\frac{x}{\xi(T)}\right]}},
\nonumber\\
\varphi_s&=&\frac{2T I}{\pi eN_0Ds}\int_{0}^x
dx'\,\frac{dx'}{|\Delta_s(x')|^2},
\label{Deltas}
\end{eqnarray}
where $\xi(T)=\sqrt{\pi D/4(T_{C}-T)}$ is the superconducting coherence length
in the vicinity of $T_{C}$.

The free energy barrier $\delta F_{+2\pi}$ in Eq. (\ref{Gamma111}) is set
by the difference
\begin{align}
\delta F_{+}= & F[\Delta_{s}(x)]-F[\Delta_{m}(x)]=\delta
F\bigg[\sqrt{2\cos\alpha-1}
\nonumber\\
& - \sqrt{\frac{2}{3}}\frac{I}{I_C} \arctan\bigg(\frac{\sqrt{3}}{2}\sqrt{\frac{2\cos\alpha-1}{1-\cos\alpha}}\bigg)\bigg],
\label{dF}
\end{align}
where $\delta F$ is defined in Eq. (\ref{dF0}). The free energy
barrier $\delta F_{-}$ for ``negative'' TAPS is determined
analogously and is related to $\delta F_{+}$ as follows
\begin{eqnarray}
\delta F_{-}=\delta F_{+} + \frac{\pi I}{e}.
\end{eqnarray}

\subsection{Pre-exponent}

Now let us turn to the pre-exponent $B_{\pm 2\pi}$ in the
expression for the TAPS rate (\ref{Gamma111}). For simplicity we
first analyze the TAPS rate in the zero current limit in which
case $\delta F_{+}=\delta F_{-}=\delta F$ and $B_{+}=B_{-}=B$. In
order to evaluate $B$ one should go beyond the stationary free
energy functional (\ref{FGL}) and include time-dependent
fluctuations of the order parameter field $\Delta(x,\tau)$. In
Ref. \onlinecite{mh} this task was accomplished within the
framework of a TDGL-based analysis. Employing TDGL equation it is
possible to re-formulate the problem in terms of the corresponding
Fokker-Planck equation \cite{Langer} which can be conveniently
solved for the problem in question. Since the important time scale
within the TDGL approach is the Ginzburg-Landau time
\begin{equation}
\tau_{GL}=\frac{\pi}{8|T_C-T|},
\label{tauGL}
\end{equation}
this time also naturally enters the expression for the pre-exponent
$B$ derived in Ref. \onlinecite{mh}.

Unfortunately the TDGL approach fails below $T_{C}$. For the sake of
illustration, let us for a moment ignore both the scalar and
the vector potentials. The
TDGL action for the wire is then usually written in the form
\begin{eqnarray}
 S_{\mathrm{TDGL}}  =  N_{0}Ts\sum_{\omega_{n}}
\int dx \frac{\pi|\omega_{n}|}{8T}|\Delta|^{2}
+ \int d\tau F[\Delta (x, \tau )],
\label{Ssgl}
\end{eqnarray}
where the GL free energy functional  $F[\Delta (x, \tau )]$ is defined in Eq. (\ref{FGL}).
This form can be obtained from the action (\ref{a105}) by formally expanding
the kernel $\chi_{\Delta}$ in Matsubara frequencies and wave vectors
$\omega_{\mu},Dq^{2} \ll4\pi T$. Note,
however, that since the validity of the GL expansion is restricted to
temperatures $T\sim T_{C}$, the Matsubara frequencies $|\omega_{n}|=2\pi |n|T$
cannot be much smaller than $4\pi T$ for any non-zero $n$. Hence, the expansion $\Psi(1/2+|\omega
_{n}|/4\pi T)-\Psi(1/2)\to\pi|\omega_{n}|/8T$ -- which yields TDGL action
(\ref{Ssgl}) -- is never
correct except in the stationary case $\omega_{n}=0$. Already these simple
arguments illustrate the failure of the TDGL action (\ref{Ssgl}) in the
Matsubara technique. Further problems with this TDGL approach arise in the
presence of the electromagnetic potentials $V$ and $A$. We refer the reader to
the paper \cite{ogzb} for the corresponding discussion.

In view of this problem one should employ a
more accurate effective action analysis. Since the microscopic effective
action for superconducting wires \cite{ogzb,GZ01,AGZ}
cannot be easily reduced to any Fokker-Planck-type of equation
it appears difficult to directly employ the McCumber-Halperin approach
\cite{mh} in order to evaluate the pre-exponent $B$ in the expression for the
TAPS rate (\ref{Gamma111}). For this reason, below we will proceed differently
and combine our effective action formalism with the
well known general formula for the decay rate of a metastable
state expressed via the imaginary part of the free energy. This method is
applicable provided the system is not driven far from equilibrium. For the
decay rate in the thermal activation regime one has
\cite{affleck,grabert,weiss}
\begin{equation}
\Gamma=-2\frac{T^*}{T}\mathrm{Im}F(T),
\label{ImF}
\end{equation}
where $T^*$ is an effective crossover temperature between the
activation regime and that of quantum tunneling under the
potential barrier. Formally $T^*$ is defined as temperature at
which a non-trivial saddle-point solution $\Delta(\tau,x)$
describing QPS first appears upon lowering $T$. Within the
accuracy of our calculation it is sufficient to estimate $T^*$
simply by setting the QPS action $S_{QPS}(T)$ equal to activation
exponent, i.e.
\begin{equation}
S_{QPS}(T^*) \simeq \delta F(T^*)/T^*
\label{crs}
\end{equation}
At sufficiently small currents one has \cite{GZ01}
\begin{equation}
S_{QPS}(T)=A sN_0\sqrt{N_0\Delta_0(T)},
\end{equation}
where $A$ is a numerical constant of order one \cite{AGZ}. Hence,
the condition (\ref{crs}) yields $\Delta_0(T^*)\sim T^*$ or,
equivalently, $T^* =a T_C$, where the numerical factor $a <1$ is
sufficiently close to unity, i.e. $T^* \sim T_C$. As the whole
concept of TAPS is only valid at $T$ close to $T_C$, one always
has $T^*/T \sim 1$. Thus, with the same accuracy one can actually
use the expression for the decay rate in the quantum regime
\cite{weiss,sz} $\Gamma=-2\mathrm{Im}F(T)$, cf. Ref.
\onlinecite{AGZ}. Here we will retain the parameter $T^*$ for the
reasons which will be clear below.

Following the standard procedure we expand the general expression
for the effective action around both solutions (\ref{Deltam}) and
(\ref{Deltas}) up to quadratic terms in both the phase $\varphi$ and
the amplitude $\delta\Delta$.
One can verify that in the limit $\Delta_0(T)\ll T$
the contributions from fluctuating
electromagnetic fields can be ignored and we obtain
\begin{equation}
S_{s/m}=F[\Delta_{s/m}]+ \delta^{2} S_{s/m},
\end{equation}
where
\begin{align}
\delta^{2} S_{m} & = \frac{sT}{2} \sum_{\omega_{n}}\int dxdx^{\prime
}
\nonumber\\ &\times\,
\big[\delta\Delta(\omega_{n},x)\chi_{\Delta}^{(m)}(|\omega
_{n}|;x-x^{\prime}) \delta\Delta(\omega_{n},x^{\prime})\nonumber\\
&  +\, \varphi(\omega_{n},x)k_{\varphi}^{(m)}(|\omega_{n}|;x-x^{\prime
})\varphi(\omega_{n},x^{\prime}) \big],\nonumber\\
\delta^{2} S_{s} & = \frac{sT}{2} \sum_{\omega_{n}}\int dxdx^{\prime
}\;
\nonumber\\ &\times\,
\big[\delta\Delta(\omega_{n},x)\chi_{\Delta}^{(s)}(|\omega
_{n}|;x,x^{\prime}) \delta\Delta(\omega_{n},x^{\prime})\nonumber\\
&  +\, \varphi(\omega_{n},x)k_{\varphi}^{(s)}(|\omega_{n}|;x,x^{\prime
})\varphi(\omega_{n},x^{\prime}) \big].
\end{align}
Here $\omega_{n}=2\pi Tn$ are Bose Matsubara frequencies. The functions $
\chi_{\Delta}^{(m)}$ and $k_{\varphi}^{(m)}$ are expressed in terms of the
kernels $\chi_{\Delta}$, $\chi_{J}$ and $\chi_{L}$ as
follows:
\begin{align}
\chi_{\Delta}^{(m)}(|\omega_{n}|;x-x^{\prime}) &
=\int\frac{dq}{2\pi}\,
e^{iq(x-x^{\prime})}\,\chi_{\Delta}(\omega_{n},q), \nonumber\\
k_{\varphi}^{(m)}(|\omega_{n}|;x-x^{\prime})  & =
\int\frac{dq}{2\pi}\, e^{iq(x-x^{\prime})}\, \nonumber\\ &\times\,
\left(  \frac{\omega_{n}^{2}}{4e^{2}}\chi _{J}(\omega_{n},q)
+\frac{q^{2}}{4m^{2}}\chi_{L}(\omega_{n},q) \right) .
\nonumber
\end{align}
The functions $\chi_{\Delta}^{(s)}$ and
$k_{\varphi}^{(s)}$ describe fluctuations around the coordinate
dependent saddle point $\Delta_{s}(x)$, and, therefore, cannot be
easily related to $\chi_{\Delta}$, $\chi_{J}$ and $\chi_{L}$.
Fortunately, the explicit form of $\chi_{\Delta}^{(s)}$ and
$k_{\varphi}^{(s)}$ is not important for us here.

The pre-exponent $B$ in eq. (\ref{Gamma111}) is obtained by integrating over
fluctuations $\delta\Delta$ in the expression for the grand partition function. One
arrives at a formally diverging expression which signals decay of a
metastable state. After a proper analytic continuation one finds the
decay rate in the form (\ref{Gamma111}) with
\begin{align}
B=-2T^*\,\mathrm{Im}\, \prod_{\omega_{n}} \sqrt{\frac{\det\chi_{\Delta
}^{(m)}(\omega_{n})\, \det k_{\varphi}^{(m)}(\omega_{n})}
{\det\chi_{\Delta}^{(s)}(\omega_{n})\, \det k_{\varphi}^{(s)}(\omega_{n})}%
}\label{B1}%
\end{align}
Here it is necessary to take an imaginary part since one of the eigenvalues of
the operator $k_\varphi^{(s)}(0)$ is negative.

The key point is to observe that at $T \sim T_{C}$ all Matsubara frequencies
$|\omega_{n}|=2\pi T|n|$ -- except for one with $n=0$ --
strongly exceed the order
parameter, $|\omega_{n}|\gg\Delta_{0}(T)$. Hence, for all such values the
function $\chi_{\Delta}(\omega_{n},q)$ approaches the asymptotic form
\cite{ogzb,GZ01}  which is not sensitive to superconductivity at all at such
values of $\omega_n$. Hence,
as long as $\Delta_0(T)\ll T$ we have
$\det\chi_{\Delta}^{(s)}(\omega_{n}) \simeq\det\chi_{\Delta}%
^{(m)}(\omega_{n}) $ and $\det k_{\varphi}^{(s)}(\omega_{n}) \simeq\det
k_{\varphi}^{(m)}(\omega_{n})$. The corresponding determinants in eq.
(\ref{B1}) cancel out and only the contribution from $\omega_{n}=0$ remains.
It yields
\begin{align}
B\simeq -2T^*\,\mathrm{Im}\, \sqrt{\frac{\det\chi_{\Delta}^{(m)}(0)\, \det
k_{\varphi}^{(m)}(0)}
{\det\chi_{\Delta}^{(s)}(0)\, \det k_{\varphi}^{(s)}(0)}}.\label{B2}%
\end{align}
The ratio of these determinants can be evaluated at zero current
with the aid of the GL free energy functional (\ref{FGL}). This
calculation yields, see Appendix A:
\begin{align}
\mathrm{Im}\, \sqrt{\frac{\det\chi_{\Delta}^{(m)}(0)\, \det k_{\varphi}^{(m)}(0)}
{\det\chi_{\Delta}^{(s)}(0)\, \det k_{\varphi}^{(s)}(0)}}
=-\frac{2\sqrt{6}}{\sqrt{\pi}}\frac{X}{\xi(T)}\sqrt{\frac{\delta F}{T}},
\label{det}
\end{align}
where, as before, $\delta F$ is defined in Eq. (\ref{dF0}).

Combining the above expressions we arrive at the final result for
the TAPS rate in the zero bias limit:
\begin{align}
\Gamma_{\pm }\equiv \Gamma =\frac{4\sqrt{6}}{\sqrt{\pi}}
T^*\frac{X}{\xi(T)}\sqrt
{\frac{\delta F}{T}} \exp\left[ -\frac{\delta F}{T}\right] .\label{Gamma222}%
\end{align}

Turning to the case of non-zero bias one can essentially repeat
the whole calculation which now yields two different TAPS rates
$\Gamma_{\pm }$. Of practical importance is the limit of transport
currents $I$ sufficiently close to the critical one, i.e.
$1-I/I_C\ll 1$. In this regime $\Gamma_{-}$ is negligibly small
whereas $\Gamma_{+}$, on the contrary, increases since the free
energy barrier
\begin{eqnarray}
\delta F_{+}(I)=\frac{4\times 6^{3/4}}{15}\delta F
\left(1-\frac{I}{I_C}\right)^{5/4} \label{dF+}
\end{eqnarray}
becomes lower than that at smaller currents. Accordingly, TAPS can be
detected easier in this limit \cite{Bezr08}.

The pre-exponent $B_{+}$ has essentially the same form as that
defined by Eq. (\ref{Gamma222}), one just needs to replace $\delta
F\to \delta F_{+}(I)$ and $T^*\to T^*(I)$. In
the limit $T_C-T\ll T_C$ considered here the current dependence of
the crossover temperature $T^*$ appears insignificant in most cases
and with sufficient accuracy one can set $T^*(I)\simeq T^*$. Indeed, very
generally one can express $T^*(I)=T^* f(I/I_C(T^*))$, where
$I_C(T^*)$ is the critical current at temperature $T^*$ and $f(x)$
is some universal function with $f (x \ll 1) \simeq 1$. Having in
mind the strong temperature dependence of $I_C(T)$ in the
temperature interval $T_C-T\ll T_C$ we find
$$
I/I_C(T^*) < I_C(T)/I_C(T^*)\sim (T_C-T)^{3/2}/T_C^{3/2}\ll 1,
$$
and, hence, $T^*(I)\approx T^*(0)\equiv T^*$. Thus, in the vicinity
of the critical current $I_C(T)-I \ll I_C(T)$ the TAPS rate can be expressed in the form
\begin{eqnarray}
\Gamma_{+}\simeq  8.84 T^*\frac{X}{\xi(T)}\sqrt {\frac{\delta
F}{T}} \left(1-\frac{I}{I_C}\right)^{5/8} 
\,e^{-\delta F_+(I)/T},
\label{IIC}
\end{eqnarray}
where $\delta F_+(I)$ is defined in Eq. (\ref{dF+}) and the
numerical prefactor is again established from the calculation of
the fluctuation determinants which is fully analogous to that
presented in Appendix A.

Summarizing all the above results and substituting $T^*=aT_C$ we
arrive at the final expression for the TAPS rates
\begin{eqnarray}
\Gamma_{\pm }(I)=\kappa
aT_C\frac{X}{\xi(T)} \sqrt{\frac{\delta
F_{\pm}(I)}{T}}\exp\left[-\frac{\delta F_{\pm}(I)}{T}\right],
\label{Gammapm}
\end{eqnarray}
where $\kappa (I)$ is a smooth function of $I$ varying from
$\kappa (0) \simeq 5.53$ to $\kappa \simeq 8.74$ at $I_C-I \ll I_C$
and, as before, the numerical prefactor $a$ is of order (and
slightly smaller than) one. Eq. (\ref{Gammapm}) is the central
result of this work. This expression is supposed to be valid at
$T_C-T \ll T_C$ and at any bias current $I<I_C$ as long as $\delta
F_{\pm }(I) \gg T$.

\subsection{Comparison with McCumber-Halperin result}

Let us compare our result (\ref{Gammapm}) with the expression for the TAPS
rates derived in Ref. \onlinecite{mh} from the TDGL-type of analysis.
We observe that Eq. (\ref{Gammapm}) does not contain the
Ginzburg-Landau time $\tau_{GL}$ and exceeds the corresponding expression
\cite{mh} by the factor $\sim T^*\tau_{GL} \sim (1-T/T_C)^{-1} \gg 1$.
On top of that, in the
vicinity of the critical current the pre-exponent in the TAPS rate (\ref{IIC})
depends on $I$ as $B_{+} \propto (1-I/I_C)^{5/8}$ in contrast to
the result  \cite{mh} $B_{TDGL} \propto  (1-I/I_C)^{15/8}$.

In order to understand the origin of these differences let us  --
just for the sake of illustration -- for a moment adopt the TDGL
action (\ref{Ssgl}) and re-calculate the TAPS rate $\Gamma_{TDGL}$
employing Eq. (\ref{ImF}). Since the whole calculation of the
fluctuation determinants remains the same (see Appendix A) we
should only re-evaluate the crossover temperature which we now
denote as $T^*_{TDGL}$. To this end we again first set $I \to 0$
and consider fluctuations of the order parameter around the saddle
point $\Delta_s(x)$ along the unstable direction (\ref{ep1})
choosing
\begin{eqnarray}
\delta\Delta(\tau,x)=i\frac{\cos\left(2\pi
T^*_{TDGL}\tau\right)}{\sqrt{2}\cosh\left(x/\xi\right)}C,
\end{eqnarray}
where $C$ is a constant. Substituting this expression into the
linearized TDGL equation and formally treating $\tau_{GL}$ as an
independent parameter, we define the classical-to-quantum
crossover temperature $T_{TDGL}^*$ as that at which a non-zero
solution ($C\not=0$) first appears. This definition yields
\begin{equation}
T_{TDGL}^*=1/4\pi \tau_{GL} \label{tdgl1}
\end{equation}
Substituting (\ref{tdgl1}) into Eq. (\ref{Gamma222}) we arrive at the
expression for $\Gamma_{TDGL}$ just 2 times bigger than that
derived in Ref. \onlinecite{mh} in the limit $I \to 0$.

An analogous -- though slightly more complicated -- analysis can be
performed also at non-zero bias current $I$. This analysis yields
\begin{equation}
T^*_{TDGL}(I)\sim T_{TDGL}^*\left(1-\frac{I}{I_C}\right)^{5/4}.
\label{tdgl2}
\end{equation}
Combining Eqs. (\ref{tdgl1}), (\ref{tdgl2}) with the result (\ref{IIC}), we
arrive at the pre-exponent
$$
B_{TDGL}\sim \frac{1}{\tau_{GL}}\frac{X}{\xi(T)}\sqrt{\frac{\delta
F}{T}} \left(1-\frac{I}{I_C}\right)^{15/8}
$$
which is again in the agreement with Ref. \onlinecite{mh}. Thus,
with the aid of the general formula (\ref{ImF}) describing
thermally activation decay of a metastable state we confirm that
the McCumber-Halperin result \cite{mh} for the TAPS rate is
essentially correct {\it within the TDGL-type of formalism}.
Unfortunately, however, the latter formalism is inaccurate by
itself. In particular, in the expression for the TAPS rate it does
not allow to correctly obtain the classical-to-quantum crossover
temperature $T^*$.

\subsection{Temperature-dependent resistance and noise}
In order to complete our analysis let us briefly address the
relation between the above TAPS rate and  physical observables,
such as, e.g., wire resistance and voltage noise. Every phase slip
event implies changing of the superconducting phase in time in
such a way that the total phase difference values along the wire
before and after this event differ by $\pm2\pi$. Since the average
voltage is linked to the time derivative of the phase by means of
the Josephson relation, $\langle
V\rangle=\langle\dot{\varphi}/2e\rangle$, for the net voltage drop
across the wire we obtain
\begin{equation}
V=\frac{\pi}{e}\left[  \Gamma_{+}(I)-\Gamma_{-}(I)\right]  ,
\end{equation}
where $\Gamma_{\pm}$ are given by Eq. (\ref{Gammapm}).
In the absence of any bias current $I\rightarrow0$ both rates
are equal $\Gamma_{\pm}=\Gamma$ and the net voltage drop $V$
vanishes. In the presence of small bias current $I\ll I_C$
we obtain
\begin{equation}
\Gamma_{\pm}(I)=\Gamma e^{\pm\pi I/2eT}.
\end{equation}
Thus, at such values of $I$ and at temperatures slightly below $T_{C}$ the
$I-V$ curve for quasi-1D superconducting wires takes a relatively simple form
\begin{equation}
V=\frac{2\pi}{e}\Gamma \,\sinh\frac{\pi I}{2eT},
\label{IVTAPS}
\end{equation}
The zero bias resistance $R(T)=(\partial V /\partial I)_{I=0}$ demonstrates exponential
dependence on temperature and the wire cross section
\begin{equation}
\frac{e^2R(T)}{2\pi} = 2\sqrt{6\pi} \frac{aT_C}{T}\frac{X}{\xi(T)}\sqrt
{\frac{\delta F}{T}}
\exp\left[-\frac{\delta F}{T}\right].
\label{TAPS R(T)}
\end{equation}

To complete our description of thermal fluctuations in superconducting wires
we point out that in addition to non-zero resistance (\ref{TAPS R(T)}) TAPS
also cause the voltage noise below $T_{C}$. Treating TAPS as independent
events one immediately concludes that they should obey Poissonian statistics.
Hence, the voltage noise power
$$
S_{V} = 2 \int dt \langle \delta V(t)\delta V(0) \rangle
$$
is given by the sum of the contributions of both ``positive'' and
``negative'' TAPS, i.e.
\begin{eqnarray}
S_V=\frac{2\pi^2}{e^2}\left[\Gamma_{+}(I)+\Gamma_{-}(I)\right].
\end{eqnarray}
At small currents $I\ll I_C$ this expression reduces to the following
simple form
\begin{align}
S_{V}=\frac{4\pi^{2}}{e^{2}}\;\Gamma \;\cosh\frac{\pi I}{2eT}.
\end{align}
Similarly to the wire resistance the voltage noise rapidly decreases as one
lowers the temperature away from $T_{C}$. Only in the vicinity of the critical
temperature this TAPS noise remains appreciable and can be detected in
experiments.

In conclusion, we have demonstrated that the rate for thermally activated
phase slips in superconducting nanowires evaluated within the microscopic
effective action analysis turns out to be parametrically bigger as compared
to the TAPS rate derived from the TDGL-type of approach. Simultaneous
measurements of both TAPS-induced resistance and noise appears to be an
efficient way for quantitative experimental analysis of thermally
activated phase slips in superconducting nanowires.

\centerline{\bf Acknowledgments}

\vspace{0.5cm}

This work was supported in part by RFBR grant 06-02-17459. D.S.G.
also acknowledges support from DFG-Center for Functional
Nanostructures (CFN).

\appendix

\section{Evaluation of fluctuation determinants}

Let us set $I=0$ and write the Ginzburg-Landau free energy in the form
\begin{eqnarray}
F=\frac{3\delta F}{4}\int\limits_{-X/2\xi}^{X/2\xi} d\eta \bigg[
\frac{u^{\prime\,2}+v^{\prime\,
2}}{2}-u^2-v^2+\frac{(u^2+v^2)^2}{2} \bigg], \label{FFFF}
\end{eqnarray}
where we introduced $\eta =x/\xi$, $u=\,{\rm
Re}\,\Delta/\Delta_0(T)$, $v=\,{\rm Im}\,\Delta/\Delta_0(T)$. In
terms of these dimensionless variables the metastable solution
(\ref{Deltam}) reads
\begin{eqnarray}
u_m(\eta )=1,\;\; v_m(\eta )=0,
\end{eqnarray}
while the saddle point solution (\ref{Deltas}) takes the form
\begin{eqnarray}
u_s(\eta )=\tanh \eta ,\;\; v_s(\eta )=0.
\end{eqnarray}

The second variation of the free energy (\ref{FFFF}) around its
saddle point with $v(y)=0$ is
\begin{eqnarray}
\delta^2F&=&\frac{3\delta F}{4}\int\limits_{-X/2\xi}^{X/2\xi}
d\eta \bigg[ \frac{\delta u^{\prime\,2}_\eta +\delta v^{\prime\,
2}_\eta }{2} -\delta u^2-\delta v^2 \nonumber\\ && +u^2(\delta
u^2+\delta v^2)+2u^2\delta u^2 \bigg].
\end{eqnarray}
Here $\delta u$ and $\delta v$ describe fluctuations of respectively the
absolute value and the phase of the order parameter. Accordingly
the operators $\chi_\Delta$ and $k_\varphi$ read
\begin{eqnarray}
\chi_\Delta^{(s)}(0)&=&\frac{3\delta
F}{4T}\left[-\frac{d^2}{d\eta^2}+4-\frac{6}{\cosh^2\eta
}\right],
\nonumber\\
\chi_\Delta^{(m)}(0)&=&\frac{3\delta
F}{4T}\left[-\frac{d^2}{d\eta^2}+4\right],
\nonumber\\
k_\varphi^{(s)}(0)&=&\frac{3\delta
F}{4T}\left[-\frac{d^2}{d\eta^2}-\frac{2}{\cosh^2\eta }\right],
\nonumber\\
k_\varphi^{(m)}(0)&=&\frac{3\delta
F}{4T}\left[-\frac{d^2}{d\eta^2}\right].
\end{eqnarray}
In order to fix the boundary conditions we note that fluctuations of
the absolute value of the order parameter in the bulk leads are negligible.
Hence, we set
\begin{eqnarray}
\delta u(-X/2)=\delta v(X/2)=0
\label{b1}
\end{eqnarray}
Likewise, since the current density vanishes in the bulk leads we
can choose
\begin{eqnarray}
\delta v'(-X/2)=\delta v'(X/2)=0.
\label{b2}
\end{eqnarray}

Let us evaluate the eigenvalues of, say, the operator
$\chi_\Delta^{(s)}(0)$. These eigenvalues
$\Lambda_n^{(s)}=(3\delta F_0/4T)\lambda_n^{(s)}$ are obtained
from the Schr\"{o}dinger equation
\begin{eqnarray}
\left[-\frac{d^2}{d\eta^2}-\frac{6}{\cosh^2\eta }\right]\delta
u=(\lambda-4)\delta u
\end{eqnarray}
with appropriate boundary conditions. The corresponding localized
solutions of this equation have the well known form \cite{LL}:
\begin{eqnarray}
\lambda_1^{(s)}=0,\;\; \delta u_1^{(s)}(\eta
)=\sqrt{\frac{3}{4}}\frac{1}{\cosh^2\eta },
\nonumber\\
\lambda_2^{(s)}=3,\;\; \delta u_2^{(s)}(\eta
)=\sqrt{\frac{3}{2}}\frac{\sinh \eta }{\cosh^2\eta }. \label{disc}
\end{eqnarray}
In order to find the eigenvalues in the continuous spectrum we
introduce transmission, $t(\lambda)$, and reflection,
$r(\lambda)$, amplitudes of the potential well \cite{LL}
\begin{eqnarray}
t(\lambda)=\frac{(1-i\sqrt{\lambda-4})(2-i\sqrt{\lambda-4})}
{(1+i\sqrt{\lambda-4})(2+i\sqrt{\lambda-4})},\;\;
r(\lambda)=0.
\end{eqnarray}
Thus, at large negative $\eta$ the wave function has the form
$\delta v(\eta )=C_1e^{i\sqrt{\lambda-4}\,\eta
}+C_2t(\lambda)e^{-i\sqrt{\lambda-4}\,\eta }$, while at large
positive $\eta$ the same wave function is $\delta v(\eta
)=C_1t(\lambda)e^{i\sqrt{\lambda-4}\,\eta
}+C_2e^{-i\sqrt{\lambda-4}\,\eta }$. Imposing the boundary
conditions (\ref{b1}) we arrive at the following equation for the
eigenvalues $\lambda_n^{(s)}$ ($n=3,4,\dots$):
\begin{eqnarray}
F_s(\lambda_n^{(s)})=0,
\label{disp}
\end{eqnarray}
where
\begin{eqnarray}
F_s(\lambda)=\frac{1}{2i}\left(t(\lambda)e^{i\sqrt{\lambda-4}\frac{X}{\xi}}
-\frac{e^{-i\sqrt{\lambda-4}\frac{X}{\xi}}}{t(\lambda)}\right).
\end{eqnarray}
In the limit $X/\xi\to\infty$ Eq. (\ref{disp}) also applies for
the discrete eigenvalues $\lambda_1^{(s)},\lambda_2^{(s)}$ (\ref{disc}).

The eigenvalues of the operator $\chi_\Delta^{(m)}(0)$ are
obtained analogously. They are defined by the equation
\begin{eqnarray}
F_m(\lambda_n^{(m)})=0,\;\; n=1,2,\dots
\end{eqnarray}
where $F_m(\lambda)=\sin\left(\sqrt{\lambda-4}\,X/\xi\right)$.
Observing that $F_{s,m}(\lambda)\propto \prod_{n=1}^\infty
(\lambda-\lambda_n^{s,m})$ and extracting the zero eigenvalue of
$\chi_\Delta^{(s)}(0)$ in a standard way, after the integration
over this zero mode we obtain
\begin{eqnarray}
\sqrt{\frac{\det\chi_\Delta^{m}(0)}{\det\chi_\Delta^{s}(0)}}&=&
\sqrt{\frac{4}{3}}\frac{X}{\xi}\sqrt{\frac{3}{4}\frac{\delta F}{2\pi T}}
\lim_{\lambda\to 0}\lim_{X\to\infty}\sqrt{\frac{\lambda F_m(\lambda)}{F_s(\lambda)}}
\nonumber\\
&=& \frac{2\sqrt{6}}{\sqrt{\pi}}\frac{X}{\xi}\sqrt{\frac{\delta F}{T}}.
\label{det1}
\end{eqnarray}

The ratio of fluctuation determinants $\det k_\varphi^{m}(0)/\det
k_\varphi^{s}(0)$ is evaluated analogously. The operator
$k_\varphi^{(s)}(0)$ has two localized eigenfunctions $\delta
v_1^{(s)}(\eta )$ and $\delta v_2^{(s)}(\eta )$ with the
eigenvalues $E_{1,2}^{(s)}=(3\delta F/4T)\epsilon_{1,2}^{(s)}$,
\begin{eqnarray}
\epsilon_1^{(s)}=-1,\;\; \delta v_1^{(s)}(\eta
)=\frac{1}{\sqrt{2}}\frac{1}{\cosh \eta }, \label{ep1}
\\
\epsilon_2^{(s)}=0,\;\; \delta v_2^{(s)}(\eta
)=\sqrt{\frac{\xi}{X}}\tanh \eta . \label{ep2}
\end{eqnarray}
The eigenvalue $\epsilon_1^{(s)}$ is negative and, as usually, it
is associated with the unstable direction in the functional space.
The ratio of the determinants is expressed as follows
\begin{eqnarray}
\sqrt{\frac{\det k_\varphi^{m}(0)}{\det k_\varphi^{s}(0)}}
=\lim_{\lambda\to 0}\sqrt{\frac{G_m(\lambda)}{G_s(\lambda)}},
\end{eqnarray}
where $G_s(\lambda)=\sin\left(\sqrt{\lambda}\,X/\xi\right)$, while
$G_s(\lambda)$ reads
\begin{eqnarray}
G_s(\lambda)=\frac{1}{2i}\left(\tilde t(\lambda)e^{i\sqrt{\lambda}\frac{X}{\xi}}
-\frac{e^{-i\sqrt{\lambda}\frac{X}{\xi}}}{\tilde t(\lambda)}\right),
\end{eqnarray}
where $\tilde t(\lambda)=(i\sqrt{\lambda}-1)/(1+i\sqrt{\lambda})$.
Thus, we get
\begin{eqnarray}
\sqrt{\frac{\det k_\varphi^{m}(0)}{\det k_\varphi^{s}(0)}}=-i.
\label{det2}
\end{eqnarray}
Combining Eqs. (\ref{det1}) and (\ref{det2}), we arrive at Eq.
(\ref{det}).


\begin{thebibliography}{99}
\bibitem{AGZ} K.Yu. Arutyunov, D.S. Golubev, and A.D. Zaikin,
Phys. Rep. (2008), to appear; arXiv: cond-mat/0805.2118.
\bibitem {Little}W.A. Little, Phys. Rev. 156 (1967) 396.
\bibitem{la} J.S. Langer and V. Ambegaokar, Phys. Rev. {\bf 164}, 498 (1967).
\bibitem {mh} D.E. McCumber and B.I. Halperin, Phys. Rev. B {\bf 1}, 1054 (1970).
\bibitem{Webb} J.E. Lukens, R.J. Warburton, and W.W. Webb,
Phys. Rev. Lett. {\bf 25}, 1180 (1970).
\bibitem {Tinkham} R.S. Newbower, M.R. Beasley, and M.
Tinkham, Phys. Rev. B {\bf 5}, 864 (1972).
\bibitem{ZGOZ} A.D. Zaikin, D.S. Golubev, A. van Otterlo, and G.T. Zimanyi,
Phys. Rev. Lett. {\bf 78}, 1552 (1997); Usp. Fiz. Nauk {\bf 168},
244 (1998) [Physics Uspekhi {\bf 42}, 226 (1998)].
\bibitem{GZ01} D.S. Golubev and A.D. Zaikin, Phys. Rev. B 64 (2001) 014504.
\bibitem {Gio} N. Giordano, Phys. Rev. Lett. {\bf 61}, 2137 (1988); Physica B {\bf 203}, 460 (1994).
\bibitem{BT} A. Bezryadin, C.N. Lau, and M. Tinkham, Nature {\bf 404}, 971 (2000).
\bibitem{Lau} C.N. Lau, N. Markovic, M. Bockrath, A. Bezryadin, and
M. Tinkham, Phys. Rev. Lett. {\bf 87}, 217003 (2001).
\bibitem{Zgirski} M. Zgirski, K.P. Riikonen, V. Tuboltsev, and
K.Yu. Arutyunov, Nano Lett. {\bf 5}, 1029 (2005); Phys. Rev. B
{\bf 77}, 054508 (2008).
\bibitem{Altomare} F. Altomare, A.M. Chang, M.R. Melloch, Y. Yong, and C.W. Tu,
Phys. Rev. Lett. {\bf 97}, 017001 (2006).
\bibitem{Bezryadin06} A.T. Bollinger, A. Rogachev and A.
Bezryadin, Europhys. Lett. {\bf 76}, 505 (2006).
\bibitem{Bezryadin08} A. Bezryadin, J. Phys.: Cond. Mat.
{\bf 20}, 043202 (2008).
\bibitem{LV} See, e.g., A.I. Larkin and A. Varlamov, {\it Theory of
    Fluctuations in Superconductors} (Clarendon, Oxford, 2005).
\bibitem{ogzb} A. van Otterlo, D.S. Golubev, A.D. Zaikin, and G. Blatter, Eur.
Phys. J. B {\bf 10}, 131 (1999).
\bibitem{LK} A. Levchenko and A. Kamenev, Phys. Rev. B {\bf 76}, 094518 (2007).
\bibitem{Langer} J.S. Langer, Phys. Rev. Lett. 21 (1968) 973.
\bibitem{affleck} I. Affleck, Phys. Rev. Lett. {\bf 46}, 388 (1981).
\bibitem{grabert} H. Grabert, P. Olschowski, and U. Weiss, Phys. Rev. B {\bf 36},
  1931 (1987).
\bibitem{weiss} U. Weiss, {\it Quantum Dissipative Systems} (World Scientific,
Singapore, 2nd Edition, 1999).
\bibitem{sz} G. Sch\"{o}n and A.D. Zaikin, Phys. Rep. {\bf 198}, 237 (1990).
\bibitem{Bezr08} M. Sahu, M.-H. Bae, A. Rogachev, D. Pekker, T.-C. Wei,
N. Shah, P.M. Golbart, and A. Bezryadin, arXiv:
cond-mat/0804.2251.
\bibitem{LL} L.D. Landau and E.M. Lifshits, {\it Quantum
mechanics} (Pergamon, Oxford, 1962).
\end{thebibliography}
\end{document}